# Oriented growth of pentacene films on vacuum-deposited polytetrafluoroethylene layers aligned by rubbing technique

Marius Prelipceanu[*,1], Otilia – Sanda Prelipceanu[1], Ovidiu-Gelu Tudose[1], Konstantin Grytsenko[1,2], Sigurd Schrader[1]

[1]University of Applied Sciences Wildau, Faculty of Engineering, Department of Engineering Physics, Friedrich-Engels-Strasse 63, D-15475 Wildau, Germany.

[2]Institute of Semiconductor Physics, 45 Nauki pr., Kyiv, 03028, Ukraine.

**Abstract**

*A new method for preparation of high quality dielectric thin films made of polytetrafluoroethylene (PTFE) is described. This method includes film formation by means of a special kind of vacuum deposition polymerization (VDP) of PTFE, assisted by electron cloud activation. Rubbing of those layers makes them orienting substrate materials which induce spontaneous ordering of deposited chromophore layers. We investigated structure and morphology of PTFE layers deposited by vacuum process in dependence on deposition parameters: deposition rate, deposition temperature, electron activation energy and activation current. Pentacene (PnC) layers deposited on top of those PTFE films are used as a tool to demonstrate the orienting ability of the PTFE layers. The molecular structure of the PTFE films was investigated by use of infrared spectroscopy. By means of ellipsometry, values of refractive index between 1.33 and 1.36 have been obtained for PTFE films in dependence on deposition conditions. Using the cold friction technique orienting PTFE layers with unidirectional grooves are obtained. On top of these PTFE films oriented PnC layers were grown. The obtained order depends both on the PTFE layer thickness and on PnC growth temperature.*

---

[*] Marius Prelipceanu, Tel: +49 3375 508 524, Fax: +49 3375 508 503, E-Mail: mariusp@igw.tfh-wildau.de

# 1. Introduction

Organic electronics, in particular, organic field effect transistors (OFET) is a fast developing field of research and technological development [1-3]. Pentacene (PnC) is one of the most extensively studied organic semiconductors for OFETs due to its relatively high carrier mobility [2]. Ordered molecular materials are used in electronic and photonic organic devices for obtaining anisotropic properties. Therefore, techniques for formation of high-quality films play an important role in the development of organic thin film devices. For such applications, uniform films with the thickness range from nanometers to submicrons are required. For electronic applications, film purity and interface characteristics influence the charge transport and energy transfer processes. For optical applications, controlling of dipole orientation is required as well as uniform thickness and low scattering loss. It is not easy to fulfill all these requirements by the wet processing. On the other hand, stable polymers like polytetrafluoroethylene (PTFE) do not dissolve in any solvent. Therefore, vacuum-based dry processing is the only possible method for deposition of such polymers. Some polymers can be evaporated by heating in vacuum, but for complex polymers low temperature plasma polymerisation should be used. Primary polymer degradation products are generated by the scission of the molecular chain at various sites and/or the cleavage of side groups or atoms. Depending on the nature of the polymer structure, the scission of polymer chains can occur either randomly or in an ordered depolymerisation mechanism. PTFE films were deposited in vacuum, but with a modified technique, which includes electron cloud activation of the decomposition products [4]. Since the discovery of the friction transfer method of PTFE hot friction transfer has been used extensively to prepare substrates materials on top of which deposited chromophores form oriented layers by self-organization [5-7]. Recently it was found that vacuum deposited and rubbed PTFE films also support growth of oriented dye layers [8-10]. Using a series of measuring techniques (e.g. ellipsometry, optical and infrared spectroscopy and atomic force microscopy) we investigated physical and optical properties of vacuum deposited PTFE and PnC thin films formed on top of these PTFE layers in order to find optimal conditions for deposition of highly oriented PnC films.

## 2. Experimental

### 2.1. Description of PTFE and PnC

PTFE is a linear polymer having the chemical structure shown in figure 1a.

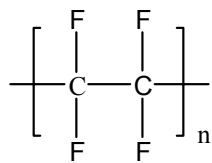   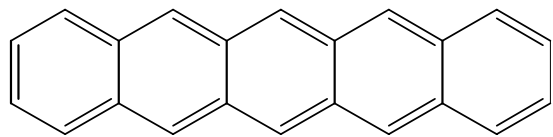

(a) polytetrafluoroethylene (PTFE)        (b) pentacene

*Fig.1. Chemical structure of: (a) polytetrafluoroethylene (PTFE), (b) pentacene (PnC).*

PTFE can be considered to be a suitable organic material serving as gate dielectric in organic field-effect transistor (OFET) devices because of its physical and chemical properties: very good chemical, photochemical and thermal stability, low dielectric constant, very low conductivity and high breakdown filed strength. PTFE is one of the most thermally stable plastic materials manifesting no appreciable decompositions below 260°C.

The chemical structure of pentacene which consists of five annulated benzene rings is shown in fig. 1b. Due to its flat conformation it can easily form crystals, which show highly anisotropic transport properties. Pentacene has a molar mass of 278.35 grams. The melting point is at about 300°C and the heat of vaporization is 74.4 kJ/mol.

### 2.2. Deposition technique

The preparation of PTFE films was carried out by use of a special vacuum deposition technique. The films were obtained by evaporation of bulk PTFE pellets in the temperature range between 300° and 450°C with electron cloud-assisted activation with typical process pressure of $10^{-2}$ Pa, an accelerating voltage of 1-3 kV and an electron activation current of 0 – 5 mA as proposed before [4,11]. The electron cloud was produced by an electron gun with a ring cathode. A computer equipped with a quartz oscillator card Sigma SQM-242 was monitoring the film thickness and deposition rate. The temperature of the crucible was monitored by a chromel-alumel thermocouple. The deposition rate depends both on the electron current used for activation and on the temperature of the crucible. At the start of a deposition run, the increase of PTFE temperature in the evaporator causes an increase of both pressure and deposition rate. Fragments are colliding with each other before reaching the substrate, losing their chemical reactivity by forming stable gaseous species, which will not be incorporated into the deposited layer on the substrate. The evaporation rate is limited by the fact that the pressure can rise only to a certain value at which a breakdown of the electrical gun occurs. Hence, there exists an operation heating temperature [4, 11], which strongly depends on the pumping speed of the vacuum system and should be determined for each installation. This method gives the possibility to have a fast control of the evaporation rate, that in general is limited by the thermal inertia of the crucible, but in this method it is controlled instantly by the electrical power, that produces the activating cloud of electrons and, therefore, changing the quantity of active species. All PTFE films were deposited at a substrate temperature of $20^0$C. Fig. 2 shows schematically the deposition installation used for the PTFE film deposition.

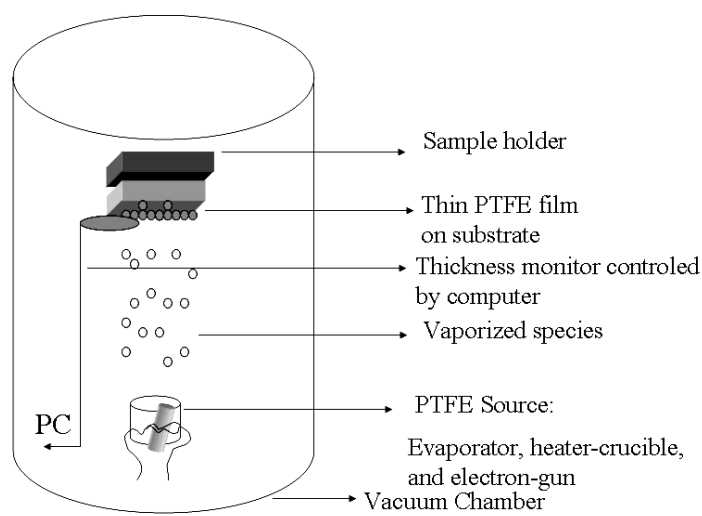

*Fig.2. Deposition set-up used for PTFE and pentacene deposition.*

PnC for fluorescence was purchased from Sigma-Aldrich and used as received. PnC films were deposited onto rubbed PTFE layers using conventional a tantalum boat heated by electric current. Important parameter which governs film formation is the temperature of the substrate. Related to this temperature, kinetic limitations such as molecular mobility, crystallization speed, and other thermodynamic factors are controlling the structure and morphology of the film. The substrate temperature was kept constant at room or elevated temperature and monitored by a chromel-alumel thermocouple. The deposition rate was chosen in the range between 0.05 and 0.2 nm/s. The distance between evaporators and substrate was 0.15 m.

### 2.3. Mechanical rubbing method

The PTFE layers deposited onto different substrates were rubbed in an unidirectional mode on a cotton surface used to clean optical systems. The cotton for friction was placed in a fixed position on an optical table. The samples were rubbed 3- 6 times on a cotton surface with a constant force and speed. The scheme for cold friction is shown in Fig. 3.

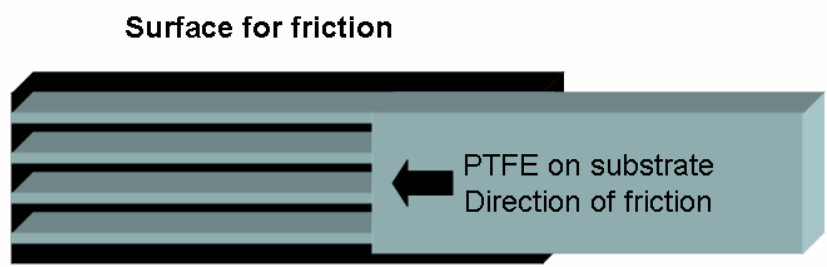

*Fig.3. Schematic representation of the cold friction technique applied to a PTFE layer.*

### 2.4. Studies of the deposited films

The surface morphology of the films was obtained using an Atomic Force Microscope (Autoprobe VP 2 Park Scientific Instruments), operating in non-contact mode in air at room temperature. The mean thickness and index of refraction ($n$) of the PTFE films were determined by means of ellipsometry using a Plasmos SD2000 Automatic Ellipsometer operating at a wavelength of 632,8 nm. The thickness of the investigated thin films was measured using a Dektak Profilometer (DEKTAK 3 from Veeco Instruments) device, which has the capability of measuring the step height down to a few nm. Polarized absorption spectra of pentacene films were obtained with a UV/VIS Spectrometer (Lambda 16 Perkin Elmer). Measurements of infrared spectra of PTFE films have been carried out by use of a Perkin-Elmer Spectrum 2000 fourier transform infrared (FT-IR) spectrometer.

**3. Results and discussions**

The analysis of the results, obtained using electron cloud assisted activation evaporation revealed that only a few important processes determine the film properties. Fig. 4 shows the influence of the evaporator temperature on the layer thickness at constant deposition time of 10 minutes. The presented curve stops well below the limiting pressure above which a decrease of deposition rate occurs due to the reason described above [4]. With increase of electron activation current the deposition rate and resulting film thickness are increasing. The maximum deposition rate obtained at a limiting pressure of 5-6 $\times 10^{-2}$ Pa was 0.18 nm/s at an activation current of 10 mA and a voltage of 3 kV. The surface relief of PTFE films deposited at different conditions onto silicon substrate is shown in Fig. 5.

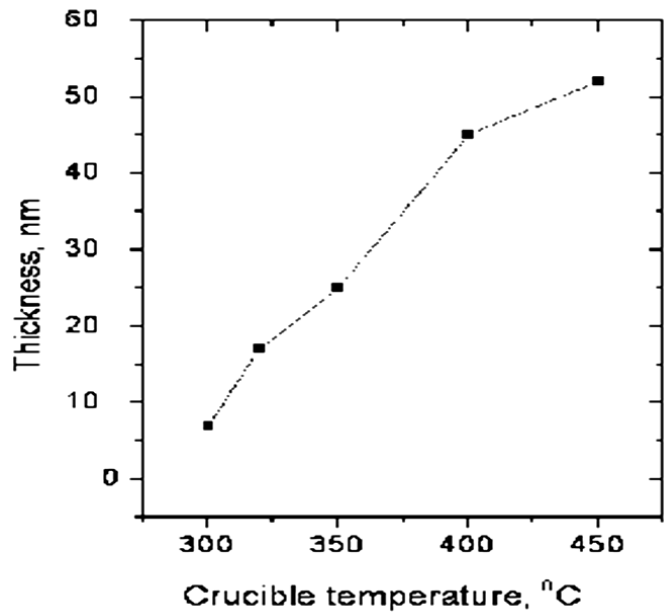

*Fig.4. Dependence of layer thickness on crucible temperature after 10 minutes of deposition, I=2mA, V=1.5 kV.*

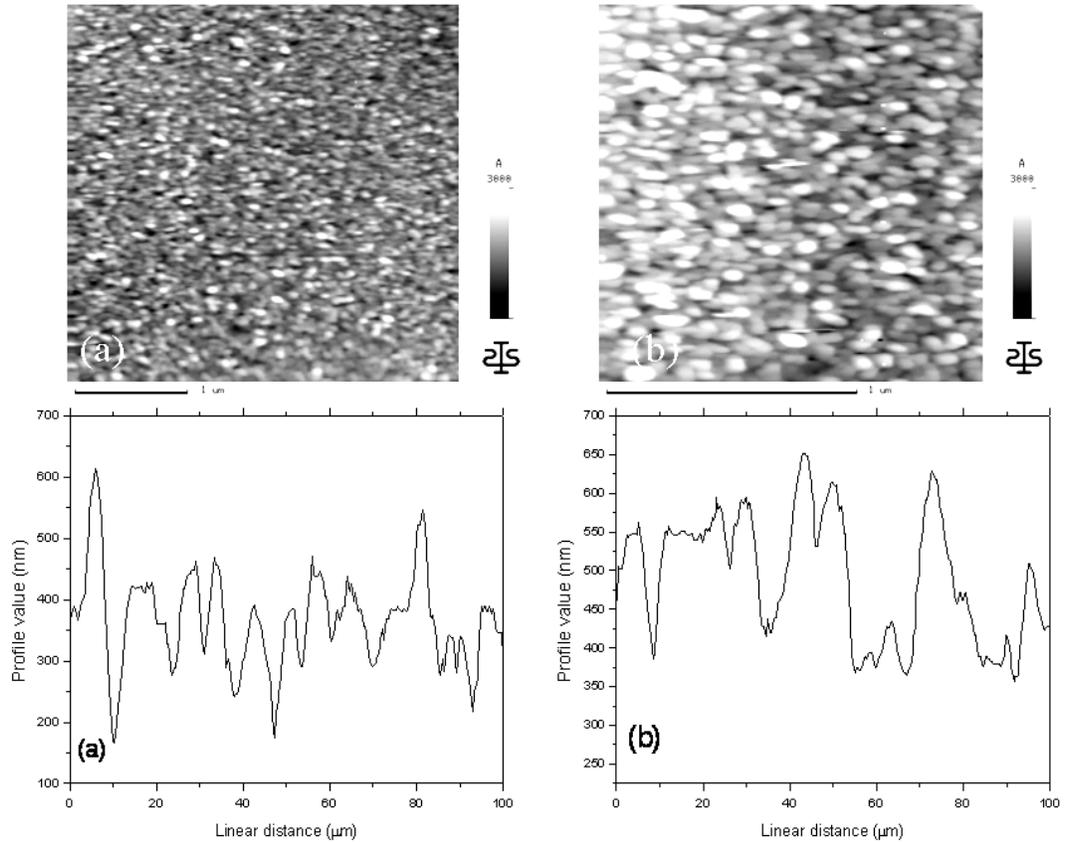

*Fig.5. Surface morphology and profile of PFTE films: (a) 2 mA and (b) 1 mA electron current activation, respectively.*

The surface of all PTFE films is smooth. For smaller electron activation current, a larger granular structure on the surface is detected. Root mean square (RMS) roughness is 1 nm and 3 nm, respectively. The obtained RMS values indicate a smoother surface occurs at higher electron activation energy. Ellipsometry results confirm the AFM investigations: thicknesses determined by both methods are comparable. In addition, a change of refractive index in dependence on electron activation energy and on current density was found, as determined by means of ellipsometry. Thus, electron activation parameters affect the surface morphology and refractive index of the PTFE films.

| Electron current, mA | 1,5 | 2 | 3 |
|---|---|---|---|
| Thickness, nm | 98 | 226 | 500 |
| Structure | amorphous | amorphous | amorphous |
| Refractive index | 1.33 | 1.34 | 1.35 |

Table 1. Refractive index of PTFE films versus activation conditions.

The IR spectra of deposited films under different activation are depicted in Fig. 6. The bands at 1161 and 1258 cm$^{-1}$ was assigned to the -CF$_2$- groups, the band at 1350 cm$^{-1}$ to groups with a double bond. The intensity of the bands at 524 and 556 cm$^{-1}$ is lower, than the intensity of the band at 736 cm$^{-1}$, thus indicating, that the material of the films is almost amorphous [11-13]. Normally at low electron activation PTFE layers are crystalline [4, 13]. An increase of electron activation current makes the films amorphous and increases the content of double bonds and side branches. Here at low activation power almost amorphous films with some double bonds but almost without branches were deposited.

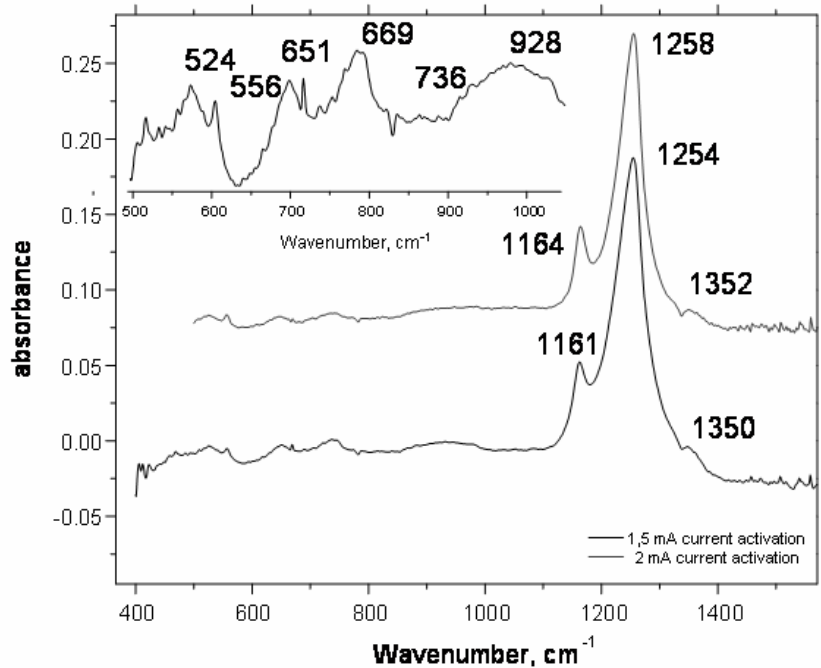

Fig.6. IR-spectra of a 500 nm-thick films, deposited by PTFE evaporation under following conditions: 1 with activation current 1,5 mA; 2 with activation current 2 mA. Inset: the magnification of IR spectra in the range from 500 to 1000 cm$^{-1}$ is shown.

After rubbing with a cotton cloth, the film surfaces were investigated by AFM and profilometry techniques. The film surface acquired ordered relief oriented in the direction of friction. Fig. 7 shows the relief and the profile of the PTFE layer after friction. The PTFE grooves have 10 -100 μm length, and about 300 nm height. Also, we can see that the spectral lines show exactly the linear structure of PTFE.

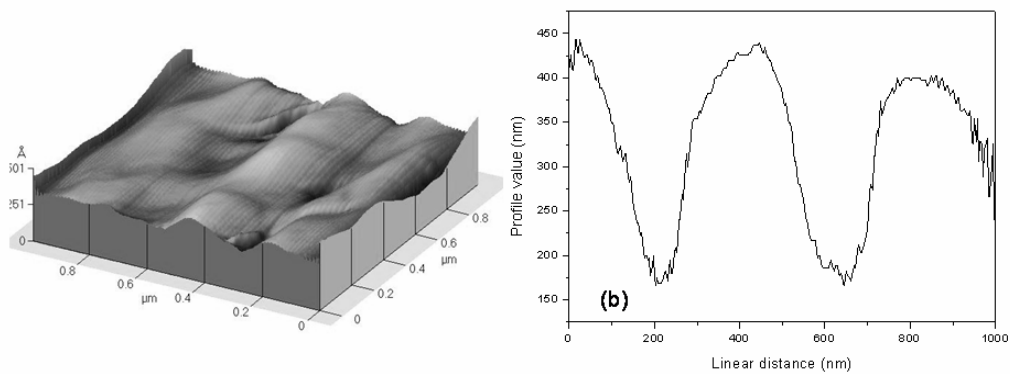

*Fig.7. The 1 μm × 1 μm AFM scan of rubbed PTFE films: (a) 3D image of the film relief; (b) profile of a series of grooves. The groove length is about 100 nm, and the height is 300 nm.*

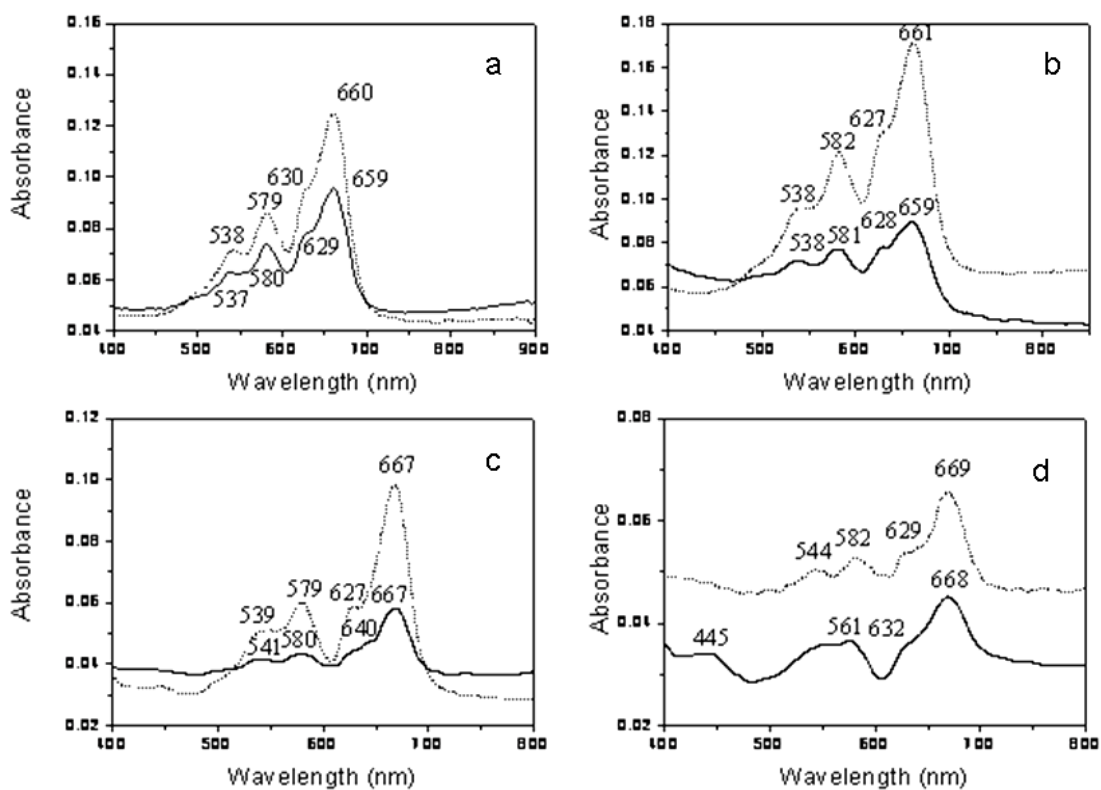

*Fig.8. Polarized absorption spectra of pentacene films for the parallel (dotted curves) and perpendicular (solid curves) orientation of the electric vector of light in*

*respect to the PTFE layer alignment. Films were deposited at the following substrate conditions: a – onto 36 nm PTFE at $20^0C$, b – onto 50 nm PTFE at $20^0C$, c – onto 90 nm PTFE at $75^0C$, d – onto 50 nm PTFE at $75^0C$. Film thicknesses by quartz monitor: a) and b) – 75 nm, c) and d) – 80 nm. Band splitting at the main absorption is 30, 34, 40 and 40 nm for a), b), c) and d) respectively.*

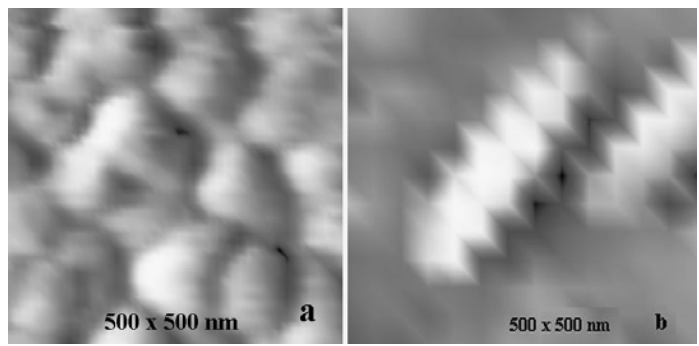

*Fig.9. Surface relief of PnC film onto rubbed PTFE sublayer.*

Measurements of electronic absorption spectra of the pentacene films deposited onto rubbed PTFE layers of different thickness have shown that orientation of the PnC films depend on both the PTFE film thickness and on substrate temperature. Optical spectra of some PnC films are presented in Fig.8. They are in a good agreement with spectra of α- and β- phase of PnC films, deposited onto both inorganic and polymer substrates, including PTFE [7, 14]. Rubbed PTFE films of about 50 nm thickness lead to the best oriented PnC films. Absorption measurements with polarized light have shown that the deposited PnC films show a pronounced dichroism. A dichroic ratio of about 2 was measured even at deposition temperature of 20°C. This dichroic ratio is larger than obtained for PnC films deposited onto friction transferred PTFE layers, and for deposition at 20°C there no dichroism was observed at all. The temperature elevation from $20^°C$ to $75^°C$ slightly enhances the PnC film orientation and changes the spectral shape. The latter two effects can be explained by the PnC molecular mobility enhancement. The former one is subject for further detailed studies. A little difference in the spectral shape indicates different molecular interactions inside of the PnC crystals dependent on deposition conditions. The crystal size and structure is also sensitive to the deposition conditions and results in modification of the absorption spectra. The optical spectra of the PnC films deposited at $75^°C$ show a small shift of all bands towards to red region and an increase of band splitting in comparison with bands of the films, deposited at $20^°C$, thus evidencing better intermolecular interactions in films, deposited at elevated temperature. Both PTFE film thickness and substrate temperature allow controlling this parameter in order to deposit PnC films with predetermined properties. The absorption of films deposited at 75°C is smaller than the absorption of films deposited at 20°C, although the quartz monitor thickness was the same for both samples. Obviously, a re-evaporation took place already at 75°C as mentioned before. By AFM no preferred crystal orientation was found in all PnC films. The typical relief of a PnC film on a PTFE aligned layer is shown in Fig.9a. The crystal size is in the range of 80 to 200 nm, depending on deposition conditions. Sometimes freely distributed needle-like PnC crystal with long axis up to 500 nm appeared (Fig.9b). Such films have low optical anisotropy. Therefore, the optical anisotropy of PnC films is due to the unidirectional arrangement of PnC molecules inside all crystals. Comparison of the obtained PnC crystals with those grown on friction-transferred PTFE layers shows that the crystals grown on the vacuum deposited, rubbed PTFE layers have smaller size and more round shape. The like effect was found for the growth of squarylium dyes on such vacuum deposited PTFE films [10]. This effect is caused by a smaller relief of the surface of the vacuum-deposited and rubbed PTFE film in comparison to the friction transferred films. In addition, some differences in the structure of friction-transferred and vacuum-deposited PTFE also plays a role. The PnC nucleation directed by PTFE edges are the main mechanism of growth of oriented PnC film as it was proposed by Brinkmann et. al.[7]. They observed that the top material domains have been enforced to grow parallel to the ledge direction due to the confinement by the PTFE nanofibrils. Only when the height of these domains exceeds that of the ledge the lateral growth of the domains is possible. The opinion about the prevailing influence of PTFE aligned on the molecular level onto dye oriented growth was expressed previously by Tanaka et. al. [8] and Wittmann et al. [5, 6, 7]. Our results seem to support the latter opinion, but the amorphous structure of our PTFE films should be taken into account. Perhaps, both mechanisms are taking place with different contributions in dependence on both the sublayer properties and deposition conditions. But even this suggestion does not explain all peculiarities, so further research should be carried out.

## 4. Conclusions

Amorphous PTFE films with RMS roughness of 1-3 nm were deposited by electron cloud-assisted deposition in vacuum. Aligned grooves and ridges on the PTFE film surface were obtained by rubbing with a cotton cloth. PTFE film thickness and growth temperature elevation influence anisotropy of pentacene film. A dichroic ratio about of 2 was obtained even when the substrate was held at room temperature. The pentacene film is oriented on the molecular level.

The strength of this technique is that the vacuum deposited, and rubbed PTFE layers have a higher orienting power than friction transferred PTFE layers so that they may favorably be used in OFETs as bottom gate dielectric which induces enhanced order in the channel material deposited on top of them. In addition, the vacuum deposited PTFE layers can also be used in top gate geometry, i.e. by deposition on top of OFET channel materials on plastic substrates.

## 5. Acknowledgements

The authors would like to thanks to Dagmar Stabenow (University of Potsdam), Ramakrishna Velagapudi (University of Applied Sciences Wildau) for the AFM and optical measurements and Dr. Oleg Dimitiriev (Institute of Semiconductor Physics, Kyiv) for the fruitful discussions. Financial support of the European Commission under contract number: HPRN-CT-2002-00327-RTN EUROFET and of Federal Ministry of Education and Research (BMBF) Project under no. Ukr 04/004 is gratefully acknowledged.

## 6. References

1. Daraktchlev M, von Muchlenen A, Nuesch F. New J. of Physics 2005; 7:113.

2. Mattis BA, Pei Y, Subramanian V. Appl.Phys. Lett. 2005; 86: 033113.

3. Misaki M, Ueda Y. Appl. Phys. Lett. 2005; 87:243503.

4. Gritsenko KP, Krasovsky AM. Chem. Rev. 2004; 103(9):3607.

5. Wittmann JC, Smith P. Nature 1991; 352:414.

6. Moulin JF, Brinkmann M, Thierry A, Wittmann JC. Adv. Mater.2002; 14(6):436.

7. Brinkmann M, Graff S, Straupe C, Wittmann JC. J.Phys.Chem. 2003; B107:10531.

8. Tanaka T, Honda Y, Ishitobi M. Langmuir 2002; 17:2192.

9. Gritsenko KP, Slominski Yu L, Tolmachev AI, Tanaka T, Schrader S. Proc.SPIE 2002; 4833: 482.

10. Gritsenko KP, Grinko DO, Dimitrev OP, Schrader S, Thierry A, Wittmann JC. Optical Memory and Neutral Networks 2004; N3:135.

11. Roeges NP. G. A Guide to the Complete Interpretation of Infrared Spectra of Organic Structures, Wiley: New York (1994).

12. Liang CY, Krimm S J. J. Chem. Phys. 1956; 25:563.

13. Gritsenko KP, Lantoukh GV. J. Applied Spectroscopy. 1990; 52:677.

14. Brinkmann M, Videva VS, Bieber A. J. Phys. Chem. 2004; A108:8170.

15. Ruiz R, Chouldhary D, Nickel B. Chem. Mater. 2004; 16:4497.

16. Pratontep S, Nüesch F, Zuppiroli L, Brinkmann M. Phys. Rev.2005; B 72:085211.